\documentclass[%
superscriptaddress,
 amsmath,amssymb,
 aps,
 twocolumn,
prb,
floatfix,
citeautoscript,
10pt
]{revtex4-2}

\setcitestyle{super}

\usepackage[caption=false]{subfig}

\usepackage{graphicx}% Include figure files
\usepackage{dcolumn}% Align table columns on decimal point
\usepackage{bm}% bold math
\usepackage{longtable}
\usepackage{xurl} %Allow URLs to break multiple lines
\usepackage{hyperref} % add hypertext capabilities
\usepackage[version=3]{mhchem}
\usepackage{color}

\newcolumntype{C}[1]{>{\centering\arraybackslash}m{#1}}
\usepackage[normalem]{ulem}
\usepackage{physics}
\usepackage{enumitem}
\usepackage{booktabs}
\usepackage{siunitx}
\usepackage{comment}

\begin{document}

\preprint{APS/123-QED}

\title{Demonstration of Parallel Multi-QPU Execution for Fragment-Based Quantum Chemistry Using On-Premises Hardware}

\author{Nils Herrmann}
\email{nils.herrmann@quantum-brilliance.com}
\affiliation{Quantum Brilliance Pty Ltd, Canberra, Australia}

\author{Mariam Akhtar}
\affiliation{Quantum Brilliance Pty Ltd, Canberra, Australia}

\author{Luke W. Bertels}
\affiliation{Computational Sciences and Engineering Division, Oak Ridge National Laboratory, Oak Ridge, TN 37831, USA}

\author{Leigh Cameron}
\affiliation{Quantum Brilliance Pty Ltd, Canberra, Australia}

\author{Raymond Chan}
\affiliation{Quantum Brilliance Pty Ltd, Canberra, Australia}

\author{Lei Cheng}
\affiliation{Chemical Sciences Division, Oak Ridge National Laboratory, Oak Ridge, TN 37831, USA}

\author{Daniel Claudino}
\email{claudinodc@ornl.gov}
\thanks{This manuscript has been authored in part by UT-Battelle, LLC,under Contract DE-AC0500OR22725 with the U.S. Department of Energy. The United States Government retains and the publisher, by accepting the article for publication, acknowledges that the United States Government retains a nonexclusive, paid-up, irrevocable, worldwide license to publish or reproduce the published form of this manuscript, or allow others to do so, for the United States Government purposes. The Department of Energy will provide public access to these results of federally sponsored research in accordance with the DOE Public Access Plan.}
\affiliation{Computational Sciences and Engineering Division, Oak Ridge National Laboratory, Oak Ridge, TN 37831, USA}

\author{Mark de Burgh}
\affiliation{Quantum Brilliance Pty Ltd, Canberra, Australia}

\author{Simon Gemmell}
\affiliation{Quantum Brilliance Pty Ltd, Canberra, Australia}

\author{Rugang Geng}
\affiliation{Quantum Brilliance Pty Ltd, Canberra, Australia}

\author{Geoff Gillett}
\affiliation{Quantum Brilliance Pty Ltd, Canberra, Australia}

\author{Travis Humble}
\affiliation{Oak Ridge National Laboratory, Oak Ridge, TN 37831, USA}

\author{Stephan Irle}
\affiliation{Computational Sciences and Engineering Division, Oak Ridge National Laboratory, Oak Ridge, TN 37831, USA}

\author{John P.\ McMahon}
\affiliation{Quantum Brilliance Pty Ltd, Canberra, Australia}

\author{Christian Ortiz}
\affiliation{Quantum Brilliance GmbH, Stuttgart, Germany}

\author{Florian Preis}
\affiliation{Quantum Brilliance GmbH, Stuttgart, Germany}

\author{Andreas Sawadsky}
\affiliation{Quantum Brilliance Pty Ltd, Canberra, Australia}

\author{Bianca Sawyer}
\affiliation{Quantum Brilliance Pty Ltd, Canberra, Australia}

\author{Reuben Singer}
\affiliation{Quantum Brilliance Pty Ltd, Canberra, Australia}

\author{Sai Meghana Tunikipati}
\affiliation{Quantum Brilliance Pty Ltd, Canberra, Australia}

\author{Cameron Walters}
\affiliation{Quantum Brilliance Pty Ltd, Canberra, Australia}

\author{Lachlan Whichello}
\affiliation{Quantum Brilliance Pty Ltd, Canberra, Australia}

\author{Adam Zegelin}
\affiliation{Quantum Brilliance Pty Ltd, Canberra, Australia}

\author{Marcus W.\ Doherty}
\affiliation{Quantum Brilliance Pty Ltd, Canberra, Australia}

\date{\today}% It is always \today, today,

\begin{abstract}
Fragment molecular orbital (FMO) quantum chemistry offers a natural route to parallel quantum computing: large molecular calculations can be decomposed into many smaller quantum subproblems that are, in principle, executable concurrently. Whether these can provide a practical benefit on real quantum hardware, however, has remained unclear. Here, we demonstrate parallel FMO calculations on three diamond-based 'Quoll' quantum computers operating at room temperature and installed on-site at the Oak Ridge Leadership Computing Facility. Helium clusters (He$_{n}$, $n=2,4,8,10,14$) were selected as a testbed in total energy calculations using the Quantum Wave Function Sampler (QWFS)---a selected configuration interaction (SCI) approach. We compare two complementary execution modes: synchronous shot-parallel execution, in which multiple quantum processors jointly sample the QWFS circuits and their SPAM-corrected counts are combined into a single probability distribution; and asynchronous fragment-parallel execution, in which processors independently evaluate FMO fragments using device-specific optimized circuits. Both modes increase quantum sampling throughput without systematic loss of accuracy. The three-QPU shot-parallel workflow reaches a parallel efficiency of $93.60\%$, while the fragment-parallel workflow achieves a measured efficiency of $75.01\%$, increasing to $95.17\%$ when excluding the non-useful idle time associated with non-interruptible submitted jobs. In both cases, the assembled FMO--QWFS energies remain chemically accurate compared to the classical ideal. We further show that heterogeneous device noise can affect parameter transfer in shot-parallel sampling, but that device-specific or jointly optimized parameter strategies restore accurate energies. These results provide a proof-of-concept demonstration that parallel quantum execution can improve the practical throughput of quantum chemistry workflows on prototype hardware, motivating the development of future HPC environments with many fault-tolerant quantum processors operating as a distributed resource for molecular and materials simulation.
\end{abstract}

\maketitle

\section{\label{sec:intro}Introduction}
Quantum computing has long been heralded as an accelerator for complex problems in areas of quantum simulation\cite{alexeev2024quantum,alexeev2025perspective,mazzola2024quantum,mcardle2020quantum,sajjan2022quantum,cao2019quantum,bassman2021simulating}, combinatorial optimization\cite{lubinski2024optimization,pirnay2024principle,chicano2025combinatorial,morris2024performant,sharma2025comparative,rieffel2024assessing}, machine learning\cite{schuld2021machine,cerezo2022challenges,peral2024systematic}, and various other fields. 
Among these, quantum chemistry is widely considered a promising candidate for the early achievement of quantum utility\cite{Herrmann2023} due to the combinatorial complexity of determining the exact energy spectrum within a given one-particle basis (the full configuration interaction (FCI) problem). 
This promise has led to a proliferation of algorithms targeting both noisy near-term quantum processors and future fault-tolerant architectures enabled by robust quantum error correction. 
Near-term hybrid quantum-classical algorithms, most notably the variational quantum eigensolver (VQE)\cite{Peruzzo2014,cerezo2021variational}, quantum sampling-based approaches \cite{robledo2025chemistry,yu2025quantum,Kaliakin2025}, and neural network quantum state methods\cite{sajjan2022quantum,Xia2018}, use the quantum processor to prepare and measure quantum states and distributions while classical co-processing is used for parameter optimization, post-processing, and/or subspace diagonalization. 
Prominent fault-tolerant algorithms include quantum phase estimation (QPE)\cite{kitaev1995quantum,aspuru2005simulated} and quantum Krylov methods\cite{cortes2022quantum,nandy2025quantum,takahashi2025krylov,stair2023stochastic,cortes2022fast} and require comparatively deeper circuits to implement powers of the Hamiltonian on the quantum register, necessitating error correction to maintain a reasonable signal-to-noise ratio. 

Despite these advances in the treatment of many-electron quantum systems with quantum computers, it remains unclear to what extent realistic molecular quantum systems can be treated directly on quantum hardware.\cite{alexeev2024quantum,alexeev2025perspective} 
Recent roadmaps from quantum hardware vendors project the delivery of platforms with $10^2-10^4$ logical qubits in the next 5-10 years.\cite{ibm_roadmap,iqm_roadmap,atom_roadmap,ionq_roadmap,quera_roadmap,google_roadmap,quantinuum_roadmap}
Even at these scales, many realistic molecular systems will exceed the capability of directly mapping to these architectures, especially when using non-minimal orbital basis sets which are necessary to extrapolate results to the complete basis set limit. 
Additionally, quantum chemistry workflows often require execution of many similar subproblems (multiple circuits, fragments, geometries). 
This is where classical HPC has conventionally tackled this problem with massive parallelization and algorithms, utilizing fragmentation and embedding strategies.\cite{kitaura2021fmo_review,Chung2015ONIOM,sun2020embedding,Verma2025,nusspickel2022systematic,nusspickel2023effective}
    
Interestingly, these same strategies have recently emerged as natural routes for extending quantum computing to chemically relevant systems. 
Example approaches adopted from ``classical" electronic structure theory include density matrix embedding theory (DMET)\cite{Rubin2016DMET,Otten2022,Liu2023,Rossmannek2023,Battaglia2024,Cao2023}, localized active space (LAS) methods\cite{Otten2022,Mitra2024,DCunha2023,King2025,wang2026localized},
many-body expansions (MBE)\cite{Verma2021,Liu2019Herbert,Ma2023}, embedded wave function (EWF) approaches\cite{Merz2026,shajan2026molecular,das2026quantum}, and the fragment molecular orbital (FMO) method\cite{kitaura1999fragment,mochizuki2021recent,Yamazaki2018,Lim2024FMO} (see Section~\ref{sec:relatedworks} for a more detailed discussion). 
These studies establish promising fragmentation frameworks for quantum chemistry on quantum computers. However, prior demonstrations have largely been numerical, emulator-based, or sequential in execution.\cite{Claudino2024}
In particular, to the best of our knowledge, no prior experiment has demonstrated on real quantum hardware that the natural task parallelism of FMO can reduce wall-clock time, increase throughput, or otherwise improve the practical execution of a quantum chemistry workflow.

In this work, we demonstrate an FMO-based quantum chemistry workflow parallelized via execution across multiple quantum processing units (QPUs). 
We apply our approach to the study of helium clusters of 2-14 atoms as a model for molecular clusters. 
Our analysis compares two complementary parallelization models.
We implement a synchronous, shot-parallelized execution where multiple QPUs jointly sample 1- and 2-qubit quantum circuits to estimate the ground state energy.
Furthermore, we also implement an asynchronous, fragment-parallel execution were the different FMO monomer and dimer fragment samplings are distributed over the different QPUs. 
We report the increase in sampling throughput, assess the accuracy of the evaluated ground state energies, and explore computational accuracy in the presence of heterogeneous device noise profiles. 

Our paper is organized as follows. 
Section~\ref{sec:relatedworks} surveys the relevant literature, while Section~\ref{sec:theory} outlines the FMO procedure and the hybrid quantum-classical sampling algorithm.
Section~\ref{sec:details} then describes the experimental details, highlighting the classical pre-processing, quantum circuit executions, and quantum hardware employed to generate the results. 
In Section~\ref{sec:results}, we present the results of our hardware executions, highlighting the performance gains available when distributing an FMO-based quantum workflow across a heterogeneous multi-QPU architecture. 
Finally, Section ~\ref{sec:conclusions} discusses the broader implications of our findings and outlines future prospects.

\section{\label{sec:relatedworks}Related Works}

\paragraph{Electronic-structure calculations on quantum hardware.}
Early experiments estimated the H$_2$ potential-energy curve using a photonic phase-estimation protocol\cite{Lanyon2010}, introduced the variational quantum eigensolver (VQE) by computing the HeH$^+$ potential-energy curve on a photonic processor\cite{Peruzzo2014}, and reproduced the HeH$^+$ calculation using a single NV defect in diamond.\cite{Wang2014}
Subsequently, superconducting platforms were used to compare VQE with Trotterized phase estimation for H$_2$\cite{OMalley2016} and to demonstrate VQE applied to small molecules.\cite{Kandala2017}
Beyond ground-state calculations, VQE-based studies on superconducting hardware have also addressed excited states using the quantum subspace expansion\cite{Colless2018} and quantum equation of motion methods\cite{Ollitrault2020} as well as the first quantum simulation of the Hartree--Fock approach.\cite{Google2020} It is worth noting that quantum chemistry using superconducting hardware has benefited from concurrent advances in error mitigation.~\cite{Guo2024, Weaving2025}
Alternatively, demonstrations of quantum chemistry using trapped-ion hardware have been presented for some of the same molecules in the early use cases above.\cite{Shen2017, Hempel2018}
Further improvements in scalability and accuracy of VQE in ion-traps have also been reported.\cite{Nam2020, Zhao2023, Goings2023, Zhao2024}
Trapped-ion quantum hardware has also been used for algorithms other than VQE.~\cite{Ntzel2025, Zhao2026}
Beyond instances revolving around VQE, we highlight Hamiltonian-moment expansions for hydrogen chains and H$_2$O\cite{Jones2022,Jones2024} and digital simulations of molecular NMR experiments.\cite{Kushal2023}
More recently, sample-based quantum diagonalization (SQD) and related quantum-centric approaches have used quantum-generated configurations to select classically tractable many-electron subspaces\cite{Robledo2025,yu2025quantum,Sugisaki2025,Shirakawa2025}, with applications including noncovalent interactions in molecular dimers.\cite{Kaliakin2025}
Of great relevance to the present work are the efforts by Merz et al.\cite{Merz2026}, Shajan et al.\cite{shajan2026molecular}, and Das et al.\cite{das2026quantum} in performing SQD calculations in tandem with molecular fragmentation. Wang et al. have also combined SQD calculation with LAS embedding methods.\cite{wang2026localized}

\paragraph{Embedding, fragmentation, and FMO-based quantum algorithms.}
Fragmentation and embedding are natural quantum--HPC strategies because they replace a large electronic-structure calculation with a collection of smaller, partially coupled subproblems.\cite{kitaura2021fmo_review,Chung2015ONIOM,sun2020embedding,Verma2025}
Rubin proposed density matrix embedding theory (DMET) with a quantum impurity solver\cite{Rubin2016DMET}, and subsequent studies developed related quantum-embedding formulations for molecular and materials applications.\cite{Otten2022,Liu2023,Rossmannek2023,Battaglia2024,Cao2023,Li2022}
Localized active-space methods were used to partition a correlated orbital space into local subspaces\cite{Otten2022,Mitra2024,DCunha2023,King2025}, while many-body expansions and related local-correlation approaches decompose corrections over orbital or molecular fragments.\cite{Verma2021,Liu2019Herbert,Ma2023}
Yamazaki et al. compared FMO, divide-and-conquer, and DMET-type decompositions as strategies for reducing quantum resource requirements\cite{Yamazaki2018}, and Kawashima et al. later executed a DMET-based decomposition of an H$_{10}$ ring on trapped-ion hardware.\cite{Kawashima2021}
Deep VQE\cite{Fujii2022}, divide-and-conquer UCC\cite{Yoshikawa2022}, and practical quantum-embedding formulations\cite{Li2022} similarly reduced the qubit requirements of the individual quantum subproblems.
For FMO specifically, Lim et al. introduced an FMO/VQE workflow for hydrogen clusters and demonstrated substantial reductions in the number of qubits required for each fragment calculation.\cite{Lim2024FMO}
Sugisaki, Nakano, and Mochizuki subsequently studied FMO-VQE/UCCSD identified size-consistency, orbital-invariance, and Trotterization issues in fragment-based quantum calculations.\cite{Sugisaki2024FMO}
Doi et al. provided the closest previous realization of FMO concurrency by processing LiH$_n$ fragment calculations concurrently on a GPU-accelerated quantum simulator.\cite{Doi2025}
FMO-VQE has also been applied to the ground-state energy of LiCoO$_2$, further extending the method.\cite{Choe2026}
An alternative avenue in terms of embedding, where orbital spaces are subject to partitioning rather than molecular functional groups, is found in recent downfolding approaches.\cite{Bauman2022,BAUMAN2023141,Singh2025}

\paragraph{Distributed and parallel quantum execution.}
Our parallelization approaches can be seen as variants of the ensemble model.\cite{Long2004,Stein2022}
Jones and Jacobsen assessed the status of distributed quantum computing in the pursuit of chemistry applications.\cite{Jones2024_2}
DiAdamo et al. formulated an accelerated VQE in which ansatz qubits and Hamiltonian measurements are distributed over networked quantum processors\cite{DiAdamo2021}, while Khait et al. analyzed how limited interprocessor quantum communication can improve a modular VQE relative to noncommunicating processors.\cite{Khait2023}
More recently, Lee et al. experimentally implemented a molecular VQE using two spatially separated photonic processors connected through a 3-km optical-fiber network.\cite{Lee2026}
At the circuit level, concurrent quantum circuit sampling maps multiple copies of a parameterized circuit onto available qubits to collect samples in parallel on a single processor.\cite{Resch2021CQCS}
Cattelan et al. instead partitioned variational circuits into independently executable slices and demonstrated the resulting parallel variational algorithms for combinatorial optimization using simulations and quantum hardware.\cite{Cattelan2025}
Virtual-QPU abstractions have also been used to distribute independent quantum-circuit simulations over classical HPC nodes\cite{Claudino2024}, while massively parallel quantum emulators have supported the development of large VQE workflows for quantum chemistry.\cite{Shang2023}
Most closely related to shot-parallel execution, recent shot-wise distribution frameworks divide the shots of a single circuit among multiple QPUs and merge their partial counts into a common output distribution.\cite{Bisicchia2025}
A subsequent implementation combined shot-wise distribution with circuit multiprogramming and validated the resulting workload-management strategy on physical IBM quantum processors.\cite{Bisicchia2026}

\section{\label{sec:theory}Theory}

This section introduces the two methodological components underlying the quantum chemistry workflow demonstrated in this work. Subsection~\ref{ssec:FMO} provides a concise overview of the fragment molecular orbital method, focusing on the FMO2 energy decomposition and the resulting opportunities for parallel execution, while subsection~\ref{ssec:QWFS} describes the employed Quantum Wave Function Sampler method used to evaluate the individual fragment energies on the quantum processors.

% QB QPU and vQPU
\subsection{\label{ssec:FMO} The Fragment Molecular Orbital Method}

The fragment molecular orbital (FMO) method\cite{kitaura1999fragment,mochizuki2021recent}
provides a natural framework for hybrid HPC--QPU workflows by decomposing a large molecular
system into a collection of smaller fragment calculations. In the present implementation,
the classical GAMESS code\cite{barca2020recent} performs the two-body approach of FMO, denoted as FMO2, partitioning
the molecular system into $N$ non-overlapping fragments and evaluating the total energy as
\begin{equation}
E_{\mathrm{FMO2}}=\sum_X^N E_X+\sum_{X>Y}^N(E_{XY}-E_X-E_Y),
\end{equation}
where $E_X$ and $E_{XY}$ denote the monomer and dimer energies, respectively.

A key computational advantage of this decomposition is that the correlated-energy
calculations associated with the individual monomers and dimers can be performed
independently.
Consequently, the fragment calculations constitute an embarrassingly parallel workload:
different monomer and dimer problems can be distributed across multiple computational
resources without requiring communication between them during their solution. This
structure makes FMO2 particularly well suited to heterogeneous HPC--QPU environments,
in which many comparatively small quantum calculations can be executed concurrently
across multiple QPUs before their resulting energies are combined classically to obtain
the total FMO2 energy.

\subsection{\label{ssec:QWFS} The Quantum Wavefunction Sampler}

In general, the electronic many-body wave function $\ket{\Psi}$ may be expanded as a linear combination
\begin{equation}
\ket{\Psi} = \sum_i c_i \ket{i}
\end{equation}
of many-electron basis states $\ket{i}$, such as Slater determinants or configuration state functions. Finding the optimal coefficients ${c_i}$ that variationally minimize the energy expectation value of a trial wave function $\ket{\tilde{\Psi}}$ is one of the central problems of electronic structure theory. In the quantum wave function sampling (QWFS) approach~\cite{Herrmann2025}, this problem is addressed by using a parameterized quantum circuit to sample the magnitudes of the coefficients $c_i$. More precisely, each basis state $\ket{i}$ is mapped to a suitable qubit basis state $\ket{b_i}$, and the corresponding coefficient $c_i(\vec{\theta})$ is written as
\begin{equation}
\label{eq:coeff}
c_i(\vec{\theta}) = s_i \sqrt{p_i(\vec{\theta})}
=
s_i \left|\bra{b_i} U(\vec{\theta}) \ket{0}^{\otimes n_q}\right| ,
\end{equation}
where $U(\vec{\theta})$ denotes a parameterized quantum circuit acting on $n_q$ qubits, $p_i(\vec{\theta}) = |\bra{b_i} U(\vec{\theta}) \ket{0}^{\otimes n_q}|^2$ is the probability of measuring bitstring $b_i$, and $s_i$ encodes the phase information that is not contained in the measured probabilities.

\begin{figure*}
\includegraphics[width=\textwidth]{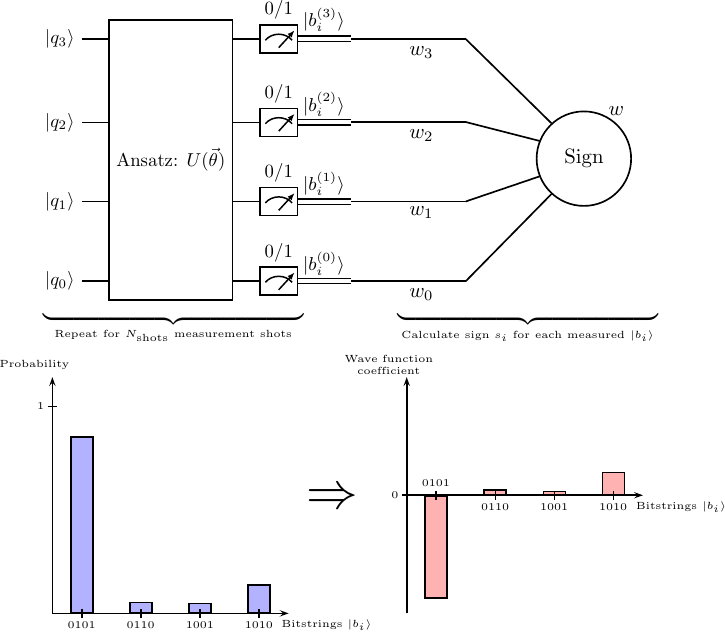}
\caption{\label{QWFS_overview.fig}Schematic overview of the QWFS workflow for an example problem size of four qubits.}
\end{figure*}

The general outline of the QWFS algorithm for a four-qubit example is illustrated in Fig.~\ref{QWFS_overview.fig}.

After a qubit encoding of the underlying quantum chemical problem has been chosen, a parameterized ansatz circuit is applied to the qubit register. The circuit prepares a superposition over computational basis states, which represent the building blocks of the electronic trial wave function $\ket{\tilde{\Psi}}$ in the chosen encoding. Measurements of the circuit produce bitstrings $\ket{b_i}$ with probabilities $p_i(\vec{\theta})$, thereby providing the coefficient magnitudes $\sqrt{p_i(\vec{\theta})}$.

The remaining phase information, represented by $s_i$, is optimized classically using a machine learning model connected to the measured qubit output. In the experiments reported here, we use a single neural network node connected to all qubit outputs, analogous to the model proposed previously for the quantum restricted Boltzmann machine~\cite{Xia2018}. For edge weights $w_k$, bias weight $w$, and a $\tanh$ activation function, the phase factor $s_i$ associated with bitstring $\ket{b_i}$ is given by
\begin{equation}
\label{eq:phase}
s_i = \tanh\left(w + \sum_{k=1}^{n_q}
\left\{
\begin{array}{lll}
+w_k && |b_i^{(k)}\rangle=|1\rangle\\
& \text{for}\\
-w_k && |b_i^{(k)}\rangle=|0\rangle
\end{array}\right.
\right)\,,
\end{equation}
where $\ket{b_i^{(k)}}$ denotes the $k$-th bit of the bitstring $\ket{b_i}$. Together, the measured probabilities and the classically obtained phase factors define the trial wavefunction coefficients via $c_i = s_i \sqrt{p_i}$.

To update the circuit and classical sign parameters, the trial energy is evaluated classically as
\begin{equation}
\label{E_calc.eq}
\langle \hat{H} \rangle
=
\frac{
\sum_{ij}
s_i s_j
\sqrt{p_i p_j}
\mel{b_i}{\hat{H}}{b_j}
}{
\sum_i s_i^2 p_i
} .
\end{equation}
The Hamiltonian matrix elements $\mel{b_i}{\hat{H}}{b_j}$ can be evaluated using the Slater--Condon rules from the one- and two-electron integrals obtained in a standard mean field calculation, typically Hartree--Fock. Once the energy has been computed, standard gradient-based or gradient-free optimization techniques can be used to update both the quantum circuit parameters and the classical sign model parameters. The resulting iterative procedure is naturally implementable on hybrid quantum classical architectures.

This sampling-based formulation of the QWFS method is particularly attractive for small scale quantum computing prototypes, where circuit throughput, calibration overhead, and hardware stability can be limiting factors. Unlike standard VQE measurement strategies, which typically require estimating the expectation values of multiple Pauli terms using basis rotation circuits, QWFS uses the quantum processor only to sample the output distribution of a single parameterized circuit. The Hamiltonian expectation value $\langle \hat{H} \rangle$ is then evaluated classically from the sampled coefficients $c_i(\vec{\theta})$ and the Hamiltonian matrix elements $\langle b_i|\hat{H}|b_j\rangle$. Consequently, the quantum circuit overhead is reduced to repeated measurements of one circuit, while the shot requirement is governed by the accuracy with which the probabilities of the relevant many-electron basis states must be resolved, rather than by the precision requirements of individual Pauli expectation value estimates. In this sense, QWFS can be more tolerant of slow or noisy prototype hardware, provided that the relevant output probabilities can be estimated with sufficient statistical confidence.

\section{\label{sec:details}Computational Details}
\subsection{\label{ssec:classical_details}Classical Computational Details}

To identify low-energy isomers, multiple initial geometries were constructed for each cluster size. These initial structures were generated by varying the relative positions of the helium atoms, including those published previously.\cite{DehdashtiJahromi2015} Full geometry optimizations at the CAM-B3LYP/6-31+G(d) level of theory were carried out for all initial configurations using the Gaussian 16 software package,\cite{g16} and the structure with the lowest electronic energy was selected as the ground-state geometry. All optimized structures were characterized as local minima through vibrational normal mode analysis.

The classical FMO2 pre-processing and automatic generation of the fragments were performed using the GAMESS program\cite{GAMESS2020}, e.g., the He$_4$ cluster was partitioned into four atomic fragments (NFRAG=4, NACUT=1), with each helium atom treated as an individual fragment. All calculations employed the Hartree--Fock method with the cc-pVDZ basis set and the FMO2 formalism. No electrostatic approximations or cutoffs were employed, such that all monomer and dimer calculations were performed in the full FMO electrostatic embedding. FCIDUMP files (as first defined by Knowles and Handy\cite{Knowles1989}) were generated and labeled using a minor modification to the GAMESS source code by invoking the FCIDUMP property driver from within the FMO module, following the FMO self-consistent field calculation of the monomers and each dimer, respectively. 

For every resulting monomer and dimer, a classical selected configuration interaction (SCI) calculation was performed to provide an ideal FMO2-SCI reference energy. These do not span the full FCI many-body space, but instead solve a deliberately restricted SCI problem chosen to fit on the two-qubit Quoll hardware (explained in greater detail in subsection \ref{ssec:quantum_details}). This restricted space is used consistently for the ideal classical reference and the deployed QWFS quantum method alike. Therefore, all accuracies reported below assess the reproduction of the target SCI problem, not recovery of the full FCI correlation energy.

\subsection{\label{ssec:quantum_details}Quantum Computational Details}

Following the mean field calculations described above, the generated FCIDUMP files containing the one- and two-electron integrals were used to construct the electron correlation problems solved with the QWFS method introduced in subsection~\ref{ssec:QWFS}. For all quantum calculations, we employed a compact encoding in which selected configuration state functions were mapped directly to computational basis states $\ket{b_i}$. This encoding allowed us to make full use of the Hilbert spaces available on the deployed prototype hardware.

For monomer fragments, corresponding to individual helium atoms, the electronic problem was mapped to a one-qubit QWFS instance. The restricted many-body space consisted of the mean field reference configuration and one double substitution into the lowest virtual spatial orbital. For dimer fragments, corresponding to weakly interacting helium dimers, the full two-qubit computational space of each Quoll was used. In this case, the restricted many-body space consisted of the mean field reference configuration, two double substitutions, and one quadruple substitution into the lowest virtual spatial orbitals. The resulting basis state mappings define the configurations sampled by the QWFS circuits and the Hamiltonian matrix elements used in the subsequent classical energy evaluation.

\begin{figure}
\centering

\subfloat[Monomer circuit.\label{fig:monomer_circuit}]{%
    \includegraphics[scale=1]{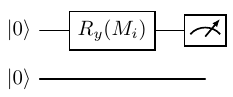}
}
\vspace{0.5em}
\subfloat[Dimer circuit.\label{fig:dimer_circuit}]{%
    \includegraphics[scale=1]{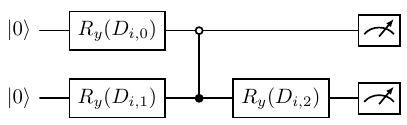}
}

\caption{
Parameterized QWFS ansatz circuits for monomers (cf.\ subfigure \ref{fig:monomer_circuit}) and dimers (cf.\ subfigure \ref{fig:dimer_circuit}) used in this work. The parameter labels $M_i$ and $D_{i,0}$, $D_{i,1}$, $D_{i,2}$ indicate device-specific optimized parameters, with $i$ denoting the device index.
}
\label{fig:qwfs_circuits}
\end{figure}

The parameterized QWFS circuits used for the monomer and dimer calculations are shown schematically in Fig.~\ref{fig:qwfs_circuits}. The monomer circuit consists of a single parameterized $R_y$ rotation acting on the first qubit, i.e., the nuclear spin of the on-site nitrogen in the NV center. The dimer circuit consists of three parameterized $R_y$ rotations together with the native anti-controlled controlled-$Z$ operation available on the Quoll hardware.

Each circuit execution was sampled using $5,000$ measurement shots. The resulting bitstring counts were converted into estimates of the probabilities $p_i$ entering the QWFS coefficient expression $c_i = s_i\sqrt{p_i}$ (Eq, \ref{eq:coeff}). The phase factors $s_i$ (Eq. \ref{eq:phase}) were evaluated using the classical sign model described in subsection~\ref{ssec:QWFS}. In all calculations reported here, this model consisted of a single neural-network node connected to all classical bit outputs. The corresponding weight and bias parameters were optimized classically for $50$ training epochs with a learning rate of $0.09$. For each epoch, the QWFS energy was evaluated from the fixed measured probability distribution and the current sign model parameters, after which automatic differentiation was used to backpropagate the energy gradient and update the sign model parameters. This optimization affects only the classical sign model and does not require additional quantum circuit executions.

The only error mitigation method applied to the hardware data was a state preparation and measurement (SPAM) correction. During the initial calibration stage, confusion matrices were obtained by preparing each computational basis state and measuring the probability of observing all possible output states. For a two-qubit device, this corresponds to preparing the four basis states $\ket{00}$, $\ket{01}$, $\ket{10}$, and $\ket{11}$. For each device, the confusion matrix was estimated using 10,000 measurement shots per prepared basis state. The resulting experimentally measured confusion matrices for Quoll-1, Quoll-2, and Quoll-3 are given in Table~\ref{tab:spam_matrices}. The resulting confusion matrix $C$ was defined such that
\begin{equation}
C_{ij} = P(\text{measured } j \mid \text{prepared } i).
\end{equation}
For an arbitrary measured probability distribution, the corresponding SPAM-corrected distribution was then obtained by maximum likelihood estimation. Specifically, if $\mathbf{p}$ denotes the unknown corrected probability vector and $\mathbf{p}^{\mathrm{meas}}$ the observed probability vector, the measurement process is modeled as
\begin{equation}
\mathbf{p}^{\mathrm{meas}} = M \mathbf{p},
\end{equation}
where $M=C^\mathrm{T}$ under the convention above. The corrected distribution was obtained by minimizing the negative log-likelihood
\begin{equation}
\mathcal{L}(\mathbf{p})
=
-\sum_i p_i^{\mathrm{meas}}
\log\left[(M\mathbf{p})_i + \epsilon\right],
\end{equation}
subject to the constraints $p_i \geq 0$ and $\sum_i p_i = 1$, where $\epsilon$ is a small numerical regularization parameter. The constrained optimization was performed using sequential least-squares programming, and the resulting normalized probability vector was used in the QWFS energy evaluation.

\begin{table}[t]
\centering
\caption{Measured SPAM confusion matrices for the three Quoll devices based on 10,000 measurements per prepared state.
Rows correspond to prepared states and columns to measured states,
ordered as $\{\ket{00},\ket{01},\ket{10},\ket{11}\}$.}
\label{tab:spam_matrices}
\begin{tabular}{c}
\textbf{Quoll-1}\\[2pt]
$\displaystyle
C_{\text{Quoll-1}} =
\begin{pmatrix}
0.7861 & 0.0309 & 0.1763 & 0.0067 \\
0.0539 & 0.7604 & 0.0687 & 0.1170 \\
0.1366 & 0.0035 & 0.8405 & 0.0194 \\
0.0385 & 0.1035 & 0.0968 & 0.7612
\end{pmatrix}$\\[16pt]

\textbf{Quoll-2}\\[2pt]
$\displaystyle
C_{\text{Quoll-2}} =
\begin{pmatrix}
0.8328 & 0.0272 & 0.1335 & 0.0065 \\
0.0723 & 0.8021 & 0.0540 & 0.0716 \\
0.0824 & 0.0079 & 0.8704 & 0.0393 \\
0.0342 & 0.0865 & 0.1047 & 0.7746
\end{pmatrix}$\\[16pt]

\textbf{Quoll-3}\\[2pt]
$\displaystyle
C_{\text{Quoll-3}} =
\begin{pmatrix}
0.7139 & 0.0370 & 0.2236 & 0.0255 \\
0.0748 & 0.6778 & 0.1129 & 0.1345 \\
0.1198 & 0.0426 & 0.7859 & 0.0517 \\
0.0993 & 0.1917 & 0.1479 & 0.5611
\end{pmatrix}$
\end{tabular}
\end{table}

\begin{figure*}
\centering
\includegraphics[width=\textwidth]{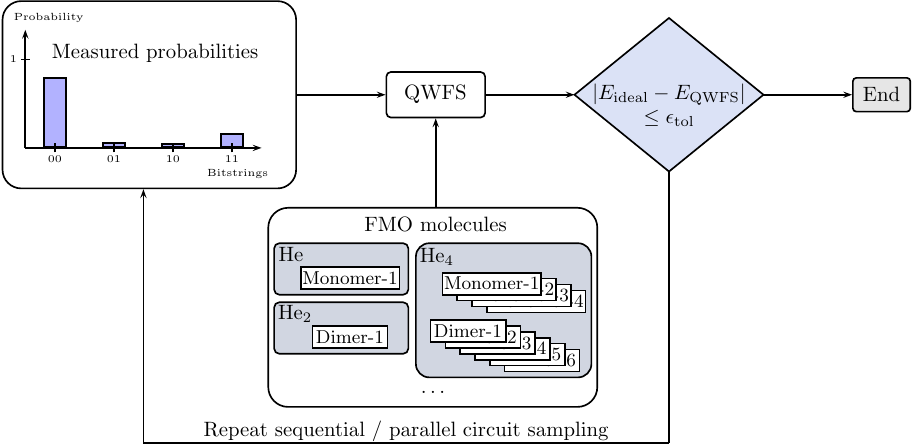}
\caption{
Schematic overview of the FMO--QWFS workflow used in this work. Following quantum circuit measurements yielding SPAM-corrected probability distributions, single point energy calculations using the QWFS method are performed. A fragment calculation is accepted if its deviation from the classical ideal energy is below the tolerance $\epsilon_{\mathrm{tol}}$; otherwise, the fragment remains in the set of fragments to be recomputed in a subsequent round of quantum circuit sampling.
}
\label{fig:fmo_qwfs_pipeline}
\end{figure*}

In principle, each FMO fragment calculation could be treated as an independent QWFS optimization, with both the quantum-circuit parameters and the classical sign-model parameters optimized separately for every monomer and dimer fragment. For the systems considered here, however, this fully iterative hardware workflow was not practical. The helium clusters studied in this work contain all monomer and dimer fragments required in the FMO2 expansion up to He$_{14}$, and optimizing the full QWFS ansatz independently for every fragment on the Quoll devices would have required an impractically long period of uninterrupted hardware access. In addition, maintaining sufficiently stable device calibrations over the corresponding optimization time would not have been realistic. We therefore used the representative optimization workflow summarized in Fig.~\ref{fig:fmo_qwfs_pipeline}.

The first stage of the workflow consisted of full representative QWFS optimizations. For each of the three Quoll devices, we optimized one representative monomer circuit and one representative dimer circuit. This produced device-specific optimized circuit parameters for the two fragment types. These optimized parameters were then kept fixed during the subsequent FMO calculations on the same device. The purpose of this step was to retain a nontrivial hardware-based QWFS workflow while avoiding the prohibitive cost of independently reoptimizing the quantum circuit for every FMO fragment.

After this representative calibration stage, the fixed-parameter circuits were executed either sequentially or in one of the parallel execution modes. The parallel execution protocols are introduced in detail in Sec.~\ref{sec:results}, directly alongside the corresponding performance and accuracy results, so that the reader can interpret the results without having to refer back to the computational details section. Independent of the execution mode, the output of the hardware stage is a SPAM-corrected probability distribution obtained from the Quoll devices.

This measured probability distribution was then reused for the QWFS ground-state energy evaluations of the required FMO fragments across the helium clusters. For each fragment, the QWFS calculation was initialized by resetting the single neural-network node used for the classical sign model, as described in Section \ref{sec:theory}. This node was retrained independently for every fragment, yielding fragment-specific wave-function phase coefficients. Together with the measured probability distribution, these phase coefficients define the full QWFS trial wave function for the fragment. The corresponding energy expectation value was then evaluated and compared with the classical ideal ground-state energy in the same restricted many-body space.

A fragment energy was accepted when the absolute deviation between the QWFS energy and the corresponding classical ideal energy was below the predefined tolerance, $\left|E_{\mathrm{ideal}} - E_{\mathrm{QWFS}}\right| \leq \epsilon_{\mathrm{tol}}$.
For all results reported in this work, we used $\epsilon_{\mathrm{tol}} = 10^{-4}$~a.u. Fragments satisfying this criterion were removed from the active set of fragments to be evaluated. Fragments that did not satisfy the criterion were retained and recomputed in a subsequent round of quantum circuit execution. The workflow was run for a maximum of 50 iterations, terminating earlier if all required monomer and dimer fragment energies satisfied the above acceptance criterion. Consecutively, the total FMO energies were assembled from the lowest obtained fragment energies.

\subsection{Quantum Hardware Specification}
\begin{figure*}
\centering
\includegraphics[width=\textwidth]{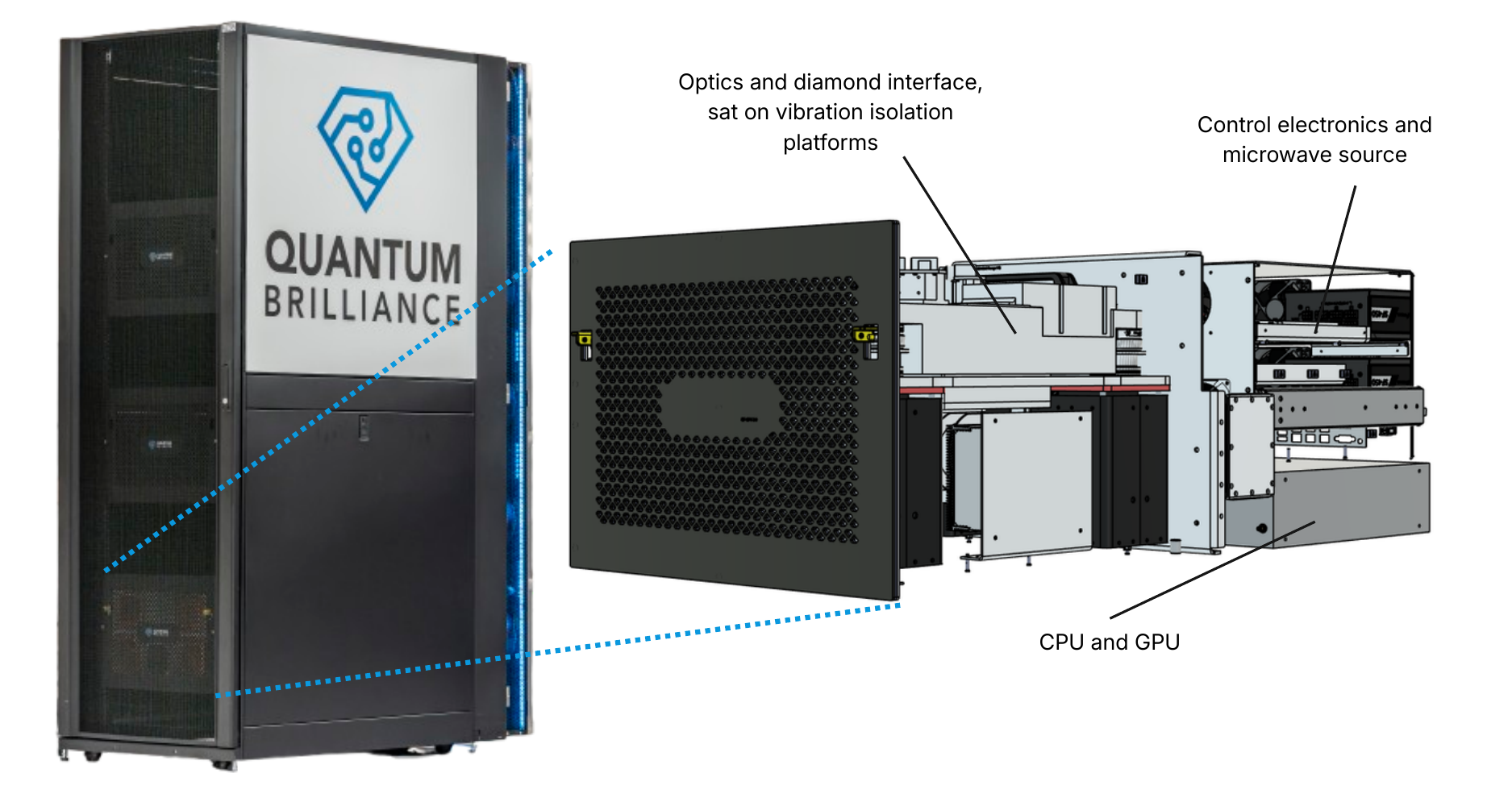}
\caption{
Installed Quoll quantum-computing cluster and internal hardware layout. Left: the fully assembled rack housing Quoll-1, Quoll-2, and Quoll-3 in a vertical configuration. Right: rendered mechanical layout of an individual Quoll system, highlighting the optics and diamond interface mounted on vibration-isolation platforms, the control electronics and microwave source, and the CPU/GPU classical compute subsystem.
}
\label{fig:quoll_schematic}
\end{figure*}

% QPU technology introduction
The quantum processing units (QPUs) used in this work are Quantum Brilliance’s (QB) integrated diamond-based quantum computer systems (cf.\ Fig. \ref{fig:quoll_schematic}). Each QPU combines an embedded classical compute layer, including CPU and GPU resources, with the quantum hardware required for optical initialization and readout, microwave and radio-frequency control, and automated device calibration. The quantum processor itself is a programmable two-qubit device based on a single nitrogen-vacancy (NV) centre in diamond with a universal native gate set. The nuclear spin states of a nitrogen-14 atom and a nearby carbon-13 atom are used as qubit 0 and qubit 1, respectively. The optical subsystem comprises a green 520\,nm laser focused onto the NV centre through a microscope objective lens, together with detection optics used for qubit-state readout. The nitrogen and carbon nuclear spins form the qubit register, while the electronic spin ($S=1$) of the NV centre serves as an ancillary degree of freedom for qubit initialization and readout and as a bus for mediating entangling two-qubit gates. In particular, the native controlled-Z gate between nuclear-spin qubits is implemented using a $2\pi$ rotation of the electronic spin conditioned on the joint nuclear-spin state.

% General QPU specs
Three such QPUs were deployed and installed at the Oak Ridge Leadership Computing Facility as a three-QPU cluster, referred to as Quoll. The systems were designed for integration into standard HPC environments and are mounted in a 19-inch server rack. Each QPU weighs 98.4\,kg and measures $339\times436\times1020$\,mm (H$\times$W$\times$D), allowing all three systems to be installed comfortably within a single rack cabinet. During normal operation, each QPU consumes approximately 400\,W of power, supplied via three IEC-C13 cables connected to standard 10\,A mains outlets. These independently power the CPU/GPU subsystem, the FPGA and control electronics, and a commercially available nano-positioning stage, respectively.

% Networking and interfacing
The Quoll system was made network-accessible from an ORNL test computing cluster, allowing it to be managed, monitored, and accessed as part of the facility’s broader computing environment. The installation addressed the usual operational requirements for HPC deployment, including operating-system compliance, power distribution, rack monitoring, management access, and remote support.\cite{ocp_qpu_2026}

% Calibration
Stable operation of the QPUs requires periodic calibration of several hardware subsystems. The position of the NV centre relative to the optical focal point can drift due to ambient vibration, humidity changes, and temperature fluctuations. This is corrected every 5 minutes using closed-loop feedback to the nano-positioning stage, maximizing the qubit-state readout signal. The frequencies of the NV’s optically detected magnetic resonance (ODMR) spectrum, as well as the output of the microwave and high-frequency delivery systems, are also sensitive to temperature fluctuations. The systems are therefore allowed to reach thermal equilibrium before operation, after which recalibration is performed every few minutes using the centre frequency of an ODMR spectrum. In addition, a laser-saturation measurement is performed every 24 hours to correct for changes in background fluorescence and long-term drift in the laser system. The calibration times and the measurement times for each QPU can vary based on the quality of the Solid Immersion Lens (SIL) milled onto the diamond, which changes the collection efficiency and the qubit readout time. For example, Quoll-2 has a collection efficiency that is almost double that of Quoll-1 leading to a higher effective sample rate. The other factor that significantly affects the calibration time and shot rate is the MW delivery system. For Quoll-3, the MW antenna required higher power to address the electronic transitions of the NV than for either Quoll-1 or Quoll-2. Consequently, additional delays were introduced into the gate sequence to minimise heating of the diamond during operation, which reduced overall gate performance and limited the number of available shots.

% Measured performance metrics
The three deployed Quolls were characterized using standard quantum process tomography measurements of the native single- and two-qubit gates and Ramsey measurements of the nuclear-spin coherence times. Across the three systems, the measured SPAM-corrected single-metric single-qubit (SMSQ) average gate fidelities were $0.95$--$0.98$, while the native anti-controlled-Z (ACZ) average gate fidelities ranged from $0.79$ to $0.96$. Detailed tomography protocols, gate-resolved fidelities, native gate durations, initialization and readout times, and measured $T_2^*$ coherence times for each Quoll are reported in section \ref{sec:SI:Hardware} of the Supporting Information.

\section{\label{sec:results}Results}

We now present the experimental results obtained by executing the FMO--QWFS workflow on the three Quoll quantum processors. In each experiment, the processors sample fixed-parameter QWFS circuits, producing probability distributions that are used in the classical QWFS energy evaluation described above. The resulting monomer and dimer fragment energies are then combined to obtain total FMO energies for helium clusters ranging from He to He$_{14}$.

The main objective of this section is to assess whether parallel quantum execution can increase sampling throughput without compromising the accuracy required for the FMO energy evaluation. We consider two complementary execution modes. In the synchronous shot-parallel mode, discussed in Sec.~\ref{results:synch}, the monomer and dimer quantum circuits are sampled across multiple quantum processors, and the resulting counts are combined into a single probability distribution before evaluating the QWFS energy. In the asynchronous fragment-parallel mode, discussed in Sec.~\ref{results:asynch}, different processors sample independently and concurrently, producing separate probability distributions used to evaluate different fragment energies in parallel. For both modes, we compare single-processor execution with two-processor and three-processor configurations, assessing accuracy and sampling throughput together.

\subsection{\label{results:synch}Synchronous shot-parallel FMO results}

\begin{figure*}
\centering
\subfloat[Shot-parallel pipeline schematic.\label{fig:shot_parallel:pipeline}]{%
    \includegraphics[width=0.85\textwidth]{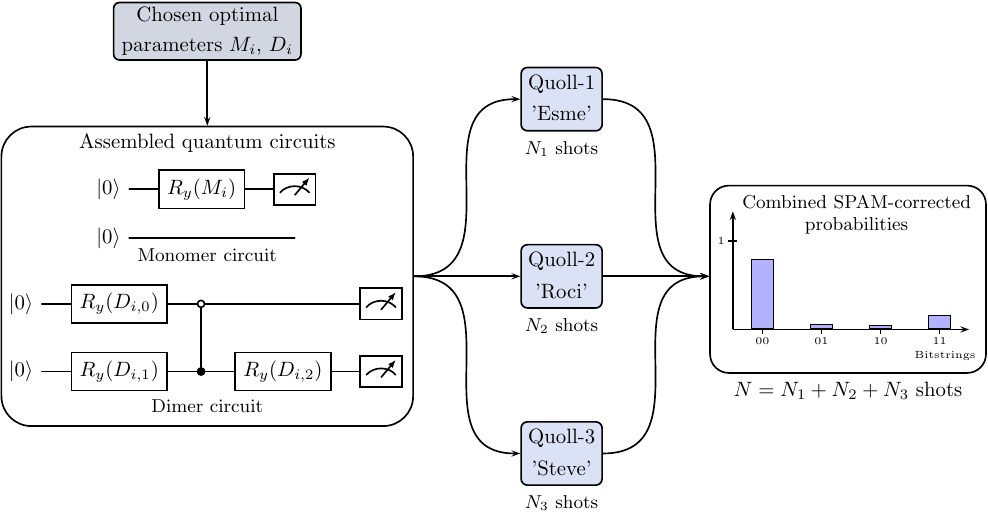}
} \\
\subfloat[Accuracy results.\label{fig:shot_parallel:accuracy}]{%
    \includegraphics[width=0.63\textwidth]{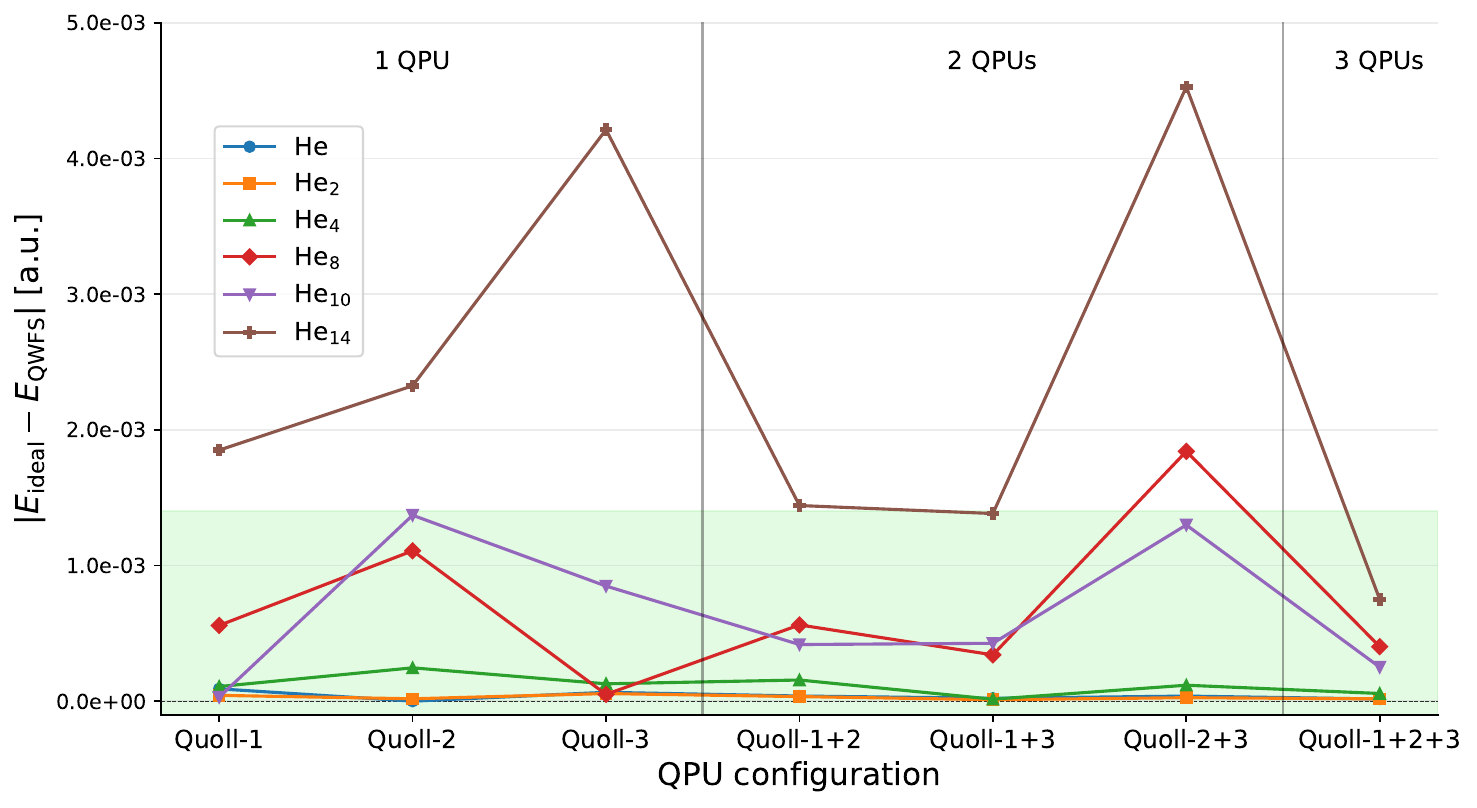}
} 
\hfill
\subfloat[Speedup results.\label{fig:shot_parallel:speedup}]{%
    \includegraphics[width=0.35\textwidth]{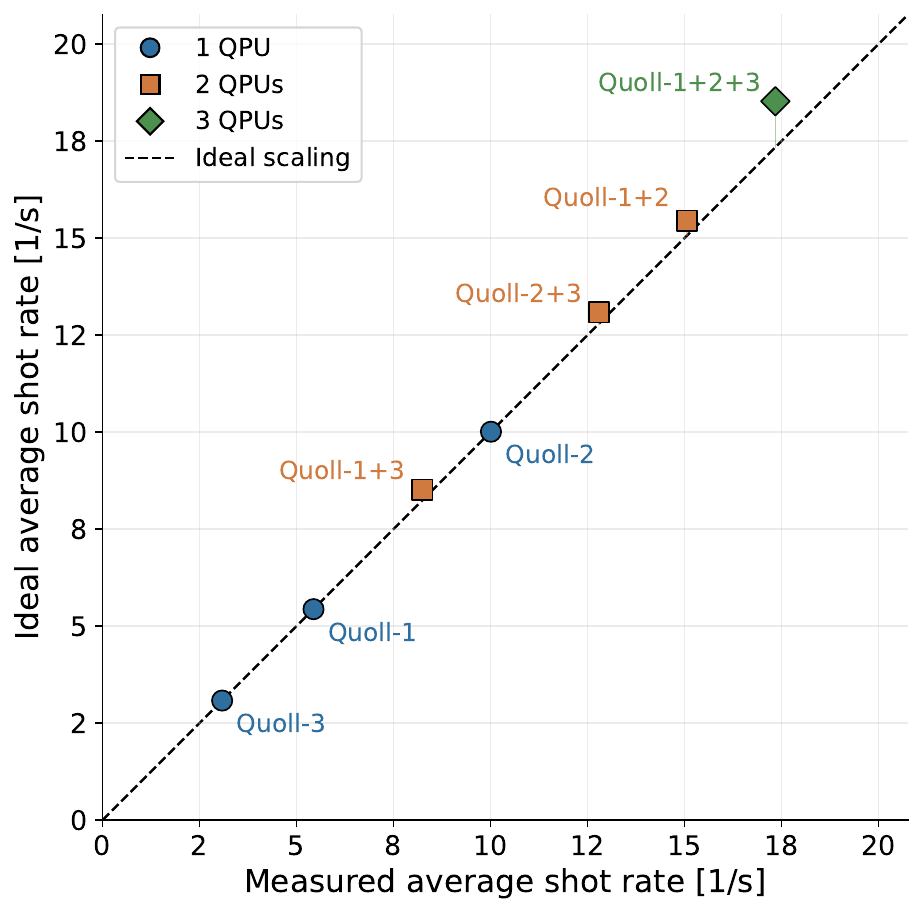}
}
\caption{\label{fig:shot_parallel}
Shot-parallel execution workflow and resulting performance.
(a) Schematic of the synchronous shot-parallel pipeline. Monomer and dimer QWFS circuits are distributed across multiple Quoll processors, and the resulting SPAM-corrected counts are combined into a single probability distribution.
(b) Absolute errors of the assembled FMO--QWFS energies for helium clusters from He to He$_{14}$. The shaded region indicates chemical accuracy ($1.4\cdot 10^{-3}$\,a.u.).
(c) Measured average shot rate compared to ideal parallel average shot rate.
}
\end{figure*}

The synchronous shot-parallel workflow and its performance are summarized in Fig.~\ref{fig:shot_parallel}.

As shown in Fig.~\ref{fig:shot_parallel:pipeline}, the shot-parallel pipeline begins from optimized QWFS parameters for the monomer and dimer fragment types. These parameters are inserted into the corresponding ansatz circuits, giving fixed-parameter circuits that are sampled during the FMO calculation. For a target total shot count $N$, the workload is distributed across the participating processors according to their independently measured average shot rates $R_i$, via
\begin{equation}
N_i = \frac{R_i}{\sum_j R_j} N\,.
\end{equation}
The shot rates $R_i$ were determined from the initial calibration measurements and kept fixed throughout the subsequent experiments. The resulting shot allocations were rounded to integer values while ensuring that the total number of assigned shots satisfied $\sum_i N_i = N$. This load-balancing strategy assigns more shots to faster processors so that all devices complete their measurements at approximately the same time. After sampling, the measured counts from each processor are SPAM-corrected and then combined into a single probability distribution. This distribution is then passed to the QWFS energy-evaluation procedure described in Sec.~\ref{ssec:quantum_details} and Fig.~\ref{fig:fmo_qwfs_pipeline}.

Figure~\ref{fig:shot_parallel:accuracy} reports the absolute deviations between the assembled FMO--QWFS energies and the corresponding classical ideal FMO energies obtained by exact diagonalization in the same restricted many-body spaces. The shaded region denotes chemical accuracy, here taken as $1.4\times 10^{-3}$~a.u. The left, middle, and right portions of the plot correspond, respectively, to single-processor execution, two-processor shot-parallel execution, and three-processor shot-parallel execution. All shot-parallel runs depicted here converged within 50 circuit-sampling iterations, with all monomer and dimer fragment energies reaching the fragment-level threshold of $10^{-4}$~a.u. relative to their classical ideal values.

Although this fragment-level threshold is more than an order of magnitude below chemical accuracy, the total FMO energy error can exceed chemical accuracy for some larger clusters. This effect is most evident for He$_{14}$ in the single-processor runs and for He$_{10}$ and He$_{14}$ in the Quoll-2+3 configuration. These larger deviations arise from the accumulation of small fragment-level errors over the increasing number of dimer contributions in the FMO expansion. Importantly, however, the data show no systematic loss of accuracy when moving from serial execution to shot-parallel execution. In fact, the three-processor configuration gives the smallest total FMO deviations across the cluster set, with all systems through He$_{14}$ lying well within chemical accuracy.

The throughput results in Fig.~\ref{fig:shot_parallel:speedup} show that the shot-parallel workflow achieves nearly ideal scaling. The measured single-processor shot rates were $5.4372$~s$^{-1}$, $10.0110$~s$^{-1}$, and $3.0815$~s$^{-1}$ for Quoll-1, Quoll-2, and Quoll-3, respectively. Despite these substantial differences in device throughput, the load-balanced two-processor configurations reached average shot rates of $15.0688$~s$^{-1}$, $8.2389$~s$^{-1}$, and $12.7949$~s$^{-1}$ for Quoll-1+2, Quoll-1+3, and Quoll-2+3, corresponding to parallel efficiencies of $97.54\%$, $96.72\%$, and $97.73\%$. The full three-processor configuration reached $17.3438$~s$^{-1}$, corresponding to a parallel efficiency of $93.60\%$.

Taken together, these results demonstrate that synchronous shot-parallel execution can increase the effective quantum sampling throughput almost linearly while preserving the accuracy of the FMO--QWFS energy estimates. This mode therefore provides a direct demonstration that multiple deployed quantum processors can be used as a combined sampling resource for a single quantum chemistry workflow.

We next analyze the three-QPU shot-parallel results in more detail by considering how the choice of fixed QWFS parameters affects the final FMO energies. The results shown in Fig.~\ref{fig:shot_parallel} correspond to one particular choice of fixed quantum circuit parameters. To examine the role of this choice more systematically, Table~\ref{tab:shot_parallel_comparison} compares several parameter selection strategies for the full three-processor configuration. 

\begin{table*}
\caption{\label{tab:shot_parallel_comparison}
Absolute deviations, in a.\ u.\ , between the three-QPU shot-parallel FMO--QWFS energies $E_{\text{QWFS}}$ and the corresponding classical ideal FMO energies $E_{\text{ideal}}$ for different choices of QWFS circuit parameters. The columns $(M_i,D_i)$ use the monomer and dimer parameters optimized on Quoll-$i$ and apply them to all three processors during shot-parallel sampling. The column $(M_1$--$M_3,D_1$--$D_3)$ uses device-specific parameters on each processor while combining the measured counts into a single probability distribution. The final column, Opt.\ $(M,D)$, uses a single monomer--dimer parameter pair optimized directly under three-QPU shot-parallel sampling.
}
\centering
\begingroup
\renewcommand{\arraystretch}{1.25} % vertical spacing
\setlength{\tabcolsep}{8pt}        % horizontal spacing; default is 6pt
\begin{tabular}{c||c|c|c|c|c|c}
\toprule
& $E_{\text{ideal}}$ & \multicolumn{5}{|c}{$|E_{\text{ideal}} - E_{\text{QWFS}}|$} \\ 
\textbf{Molecule} &  & ($M_1$, $D_1$) & ($M_2$, $D_2$) & ($M_3$, $D_3$) & ($M_1$--$M_3$, $D_1$--$D_3$) & Opt.\ ($M$, $D$) \\
\hline
He        &  -2.870\,180\,6 & 0.000\,001\,7 & 0.000\,094\,2 & 0.000\,004\,6 & 0.000\,016\,0 & 0.000\,094\,2 \\
He$_2$    &  -5.716\,974\,8 & 0.010\,858\,5 & 0.003\,397\,9 & 0.011\,775\,9 & 0.000\,017\,1 & 0.000\,039\,0 \\
He$_4$    & -11.340\,807\,9 & 0.065\,325\,4 & 0.356\,940\,9 & 0.070\,181\,5 & 0.000\,056\,6 & 0.000\,172\,3 \\
He$_8$    & -22.308\,776\,5 & 0.303\,302\,7 & 0.500\,505\,0 & 0.328\,179\,8 & 0.000\,402\,0 & 0.000\,864\,0 \\
He$_{10}$ & -28.057\,078\,2 & 0.304\,101\,6 & 0.382\,575\,6 & 0.750\,536\,8 & 0.000\,250\,7 & 0.001\,580\,6 \\
He$_{14}$ & -38.347\,086\,0 & 0.857\,049\,1 & 1.079\,087\,4 & 1.150\,230\,3 & 0.000\,750\,7 & 0.001\,257\,6 \\
\bottomrule
\end{tabular}
\endgroup
\end{table*}

In all cases, the circuits are sampled in the same three-QPU shot-parallel manner: measurements from the three processors are SPAM-corrected and combined into a single probability distribution before the QWFS energy is evaluated. The columns differ only in how the fixed QWFS parameters are chosen. In the first three columns, a single monomer--dimer parameter pair, $(M_i,D_i)$, optimized on one quantum processor, $i$, is used on all three processors. The column denoted $(M_1$--$M_3,D_1$--$D_3)$ instead assigns to each processor the parameters optimized on that same processor. Please note that this is the strategy used for the three-QPU shot-parallel results shown in Fig.~\ref{fig:shot_parallel}. Finally, the Opt.\ $(M,D)$ column uses a single new monomer--dimer parameter pair optimized directly under the full three-QPU shot-parallel sampling procedure.

Two clear trends emerge from the table: Firstly, the monomer circuits are not the limiting factor. As single-qubit circuits with one variational rotation, they reach the targeted accuracy of $10^{-4}$~a.u. in all monomer fragments for all parameter choices. In contrast, the dimer circuits seem highly sensitive to the choice of parameters and to the device on which they are sampled. Secondly, when a parameter pair optimized on one Quoll is applied uniformly across the three-processor shot-parallel configuration, none of the dimer fragments converges within 50 circuit-sampling iterations. This lack of dimer convergence leads to large accumulated FMO errors, reaching, e.g., more than $1$~a.u. for the $(M_2,D_2)$ and $(M_3,D_3)$ choices. Thus, circuit parameters that were optimized to compensate the noise characteristics of one specific quantum processor do not necessarily produce the desired probability distribution when the same circuit is sampled on other quantum processors. This effect was not observed in the two-QPU shot-parallel experiments. For all two-processor configurations, and for all tested choices of optimized monomer and dimer circuits, every monomer and dimer fragment energy converged to within the target threshold of $10^{-4}$~a.u. within at most 50 circuit sampling iterations. The breakdown seen in Table~\ref{tab:shot_parallel_comparison} therefore appears only when all three processors are combined, indicating that the sensitivity to cross-device noise heterogeneity becomes more pronounced as the number of QPUs increases.

The two remaining strategies, however, substantially mitigate this problem. When each processor uses its own locally optimized monomer and dimer parameters, $(M_1$--$M_3,D_1$--$D_3)$, the combined shot-parallel distribution yields FMO energies within chemical accuracy for all systems considered. This indicates that the processor-dependent noise can, at least for the present three-QPU experiment, be compensated by allowing the circuits sampled on different devices to differ slightly while targeting the same effective probability distribution. Similarly, the Opt.\ $(M,D)$ strategy, in which a single parameter pair is optimized directly using the combined three-QPU sampling process, also produces chemically accurate total energies for all systems except He$_{10}$, whose deviation is only slightly above the chemical-accuracy threshold.

These results highlight an important practical aspect of shot-parallel execution on NISQ hardware. The sampling throughput can be increased nearly linearly, as shown above, but the accuracy of the resulting energy estimate depends on whether the sampled circuits remain appropriate for the combined noisy device ensemble. For the present prototype room-temperature diamond processors, direct transfer of parameters optimized on one processor to a multi-processor shot-parallel setting is not sufficient for the dimer circuits. However, either device-specific parameter assignment or explicit optimization under the full shot-parallel noise environment restores chemically accurate FMO energies in all dimer fragments. We therefore interpret Table~\ref{tab:shot_parallel_comparison} as evidence that heterogeneous device noise is a central consideration in near-term distributed quantum sampling, but one that might be addressed through parameter-selection strategies tailored to the parallel execution mode.

At the same time, this conclusion should be viewed in the context of the small three-QPU prototype system studied here. The observed mitigation strategies need not necessarily scale to larger processor counts, where the distribution of device noise, calibration drift, and cross-device variability may become more difficult to compensate. Furthermore, these effects are specific to the NISQ regime of quantum computing. In the fault-tolerant setting anticipated for quantum advantage in quantum chemistry, sufficiently protected logical qubits should suppress hardware-level variability to the point where processors behave as interchangeable samplers of the same logical circuit.

\subsection{\label{results:asynch} Asynchronous fragment-parallel FMO results}

\begin{figure*}
\centering
\subfloat[Asynchronous fragment-parallel pipeline schematic.\label{fig:frag_parallel:pipeline}]{%
    \includegraphics[width=0.90\textwidth]{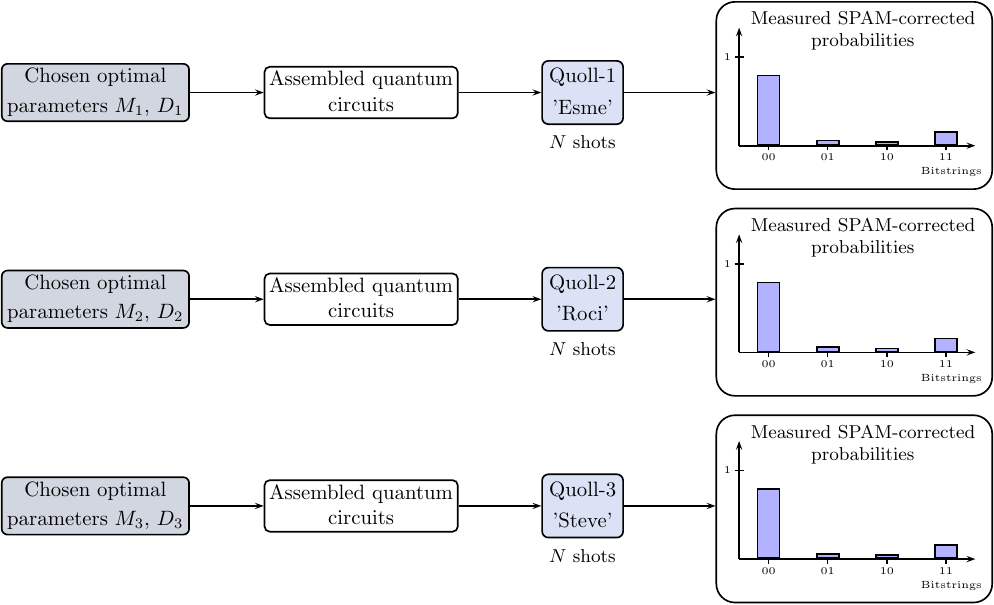}
} \\
\subfloat[Accuracy results.\label{fig:frag_parallel:accuracy}]{%
    \includegraphics[width=0.63\textwidth]{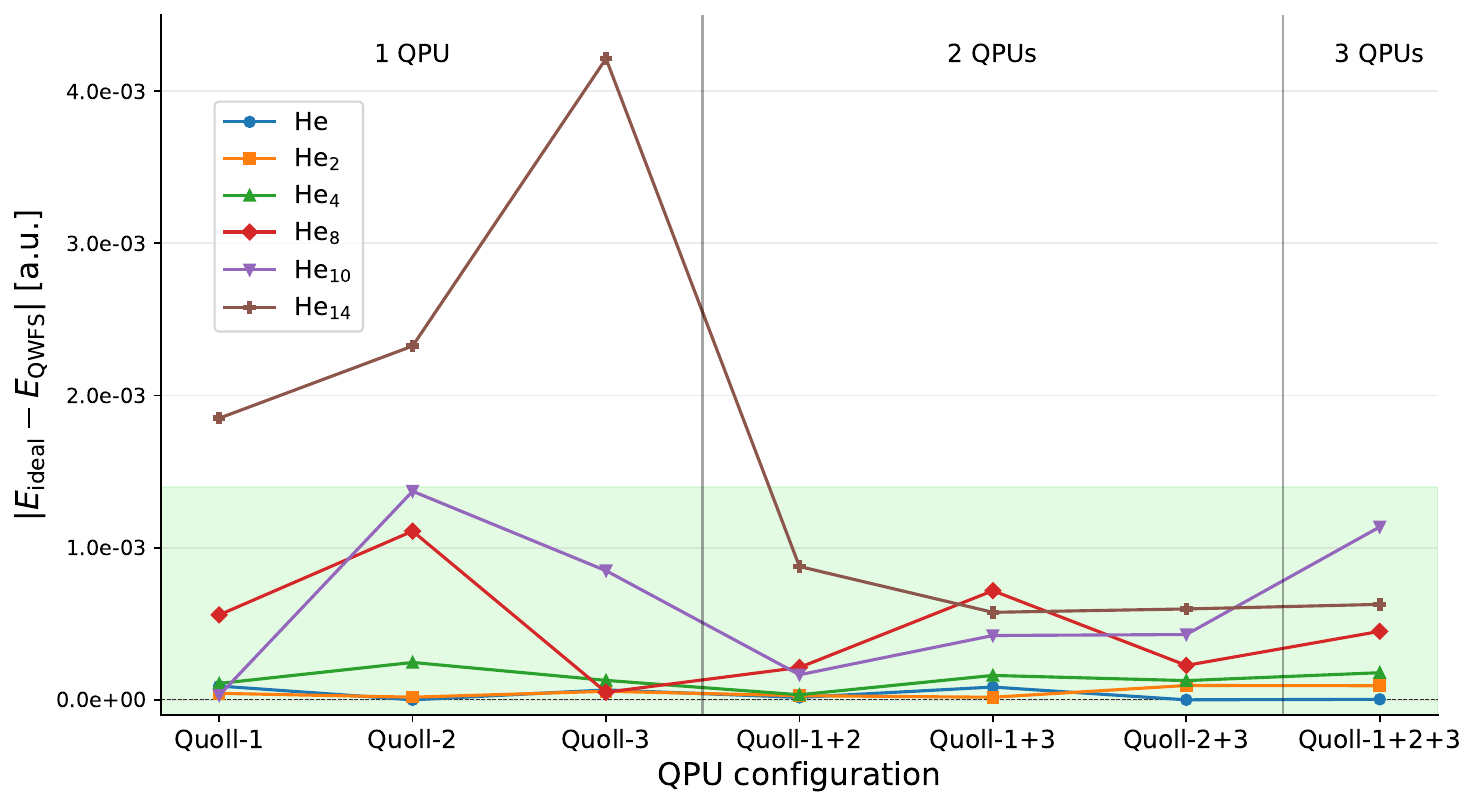}
}
\hfill
\subfloat[Speedup results.\label{fig:frag_parallel:speedup}]{%
    \includegraphics[width=0.35\textwidth]{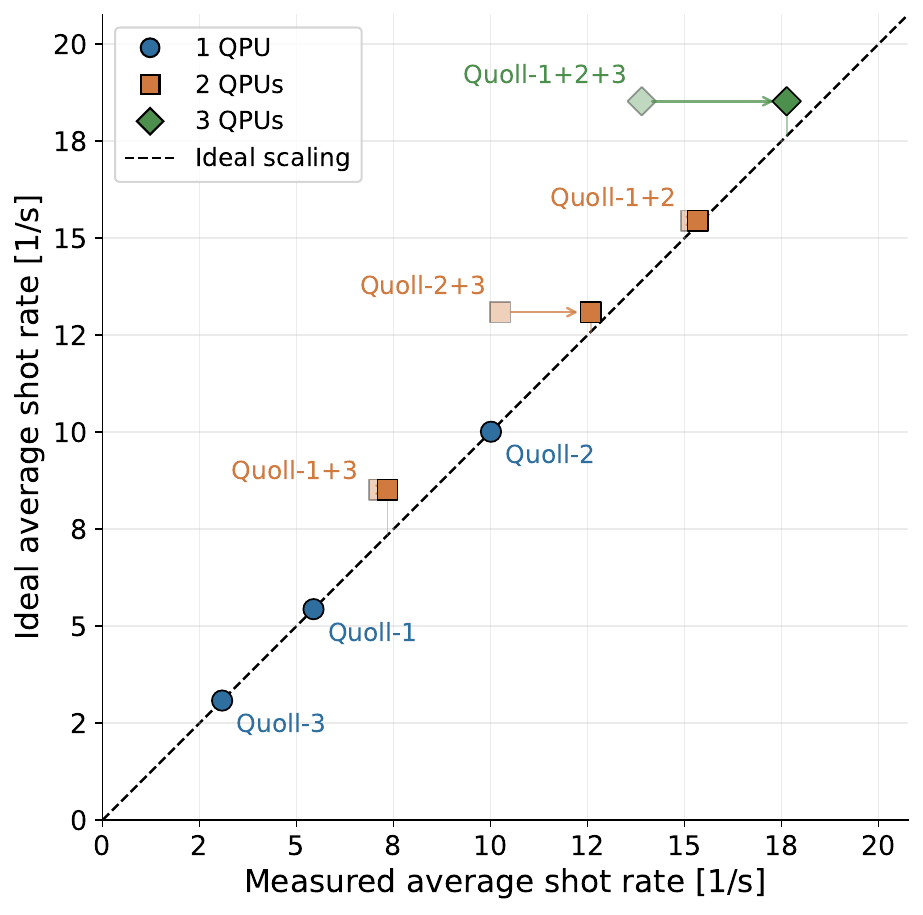}
}
\caption{\label{fig:frag_parallel} Asynchronous fragment-parallel execution workflow and resulting performance. 
(a) Schematic of the asynchronous fragment-parallel pipeline, in which independent monomer and dimer fragment evaluations are distributed across multiple QPUs. 
(b) Absolute errors of the assembled FMO--QWFS energies for helium clusters from He to He$_{14}$. The shaded region indicates chemical accuracy ($1.4\times 10^{-3}$\,a.u.).
(c) Measured average shot rates compared with the ideal parallel average shot rates inferred from the corresponding single-QPU throughputs. Transparent markers show the raw measured rates computed from the full asynchronous wall-clock time, including waiting time from non-interruptible circuit executions that continued after the relevant monomer or dimer phase had already completed. Opaque markers show idle-time-corrected rates, obtained by removing this post-completion waiting time.}
\end{figure*}

We next consider the asynchronous fragment-parallel execution mode, summarized in Fig.~\ref{fig:frag_parallel}. 

In this mode, the processors are not used to jointly sample a single circuit instance. Instead, each processor independently samples complete QWFS circuits using its own optimized monomer and dimer parameters. Thus, each processor executes the full target number of $N$ shots for its assigned circuit, produces its own SPAM-corrected probability distribution, and then sends the latter to QWFS instances to evaluate the corresponding fragment energies. Fragment calculations that reach the convergence threshold of $10^{-4}$~a.u. are removed from the queue, while the remaining fragment circuits continue to be sampled asynchronously across the available processors. The workflow is organized into two fragment phases. It first processes the monomer queue until all monomer fragments have reached the target threshold, or until the maximum number of circuit-sampling iterations has been reached, and then proceeds analogously through the dimer queue.

This execution mode avoids the parameter-selection issue observed in the three-QPU shot-parallel setting. Since each processor samples circuits using parameters optimized for that same device, the workflow does not produce a single effective probability distribution. Instead, the parallelism arises from evaluating the remaining fragment energies concurrently. The trade-off is that each circuit execution requires the full shot count on each single processor, rather than distributing the shots for one circuit across multiple processors.

The accuracy results in Fig.~\ref{fig:frag_parallel:accuracy} show that the asynchronous fragment-parallel workflow converged within 50 circuit-sampling iterations for all two-QPU configurations and for the full three-QPU configuration. As in the shot-parallel case, increasing the number of processors does not lead to a systematic loss of accuracy. In fact, for all two- and three-processor asynchronous runs, the assembled FMO--QWFS energies for all helium clusters through He$_{14}$ lie within chemical accuracy to the ideal classical FMO-SCI energies. This includes He$_{14}$, which exhibited larger deviations in some of the shot-parallel configurations.

The throughput results are shown in Fig.~\ref{fig:frag_parallel:speedup}. The transparent markers denote the raw measured average shot rates obtained from the full observed wall-clock time of the asynchronous workflow. These rates were $15.1774$~s$^{-1}$, $7.1367$~s$^{-1}$, and $10.2518$~s$^{-1}$ for Quoll-1+2, Quoll-1+3, and Quoll-2+3, corresponding to parallel efficiencies of $98.25\%$, $83.78\%$, and $78.30\%$, respectively. For the full three-QPU configuration, the raw measured average shot rate was $13.8985$~s$^{-1}$, corresponding to a parallel efficiency of $75.01\%$. These raw efficiencies are significantly lower than those observed in the shot-parallel workflow, despite the fact that the same single-processor rates enter the ideal scaling estimate.

\begin{figure*}
\centering
\includegraphics[width=0.95\textwidth]{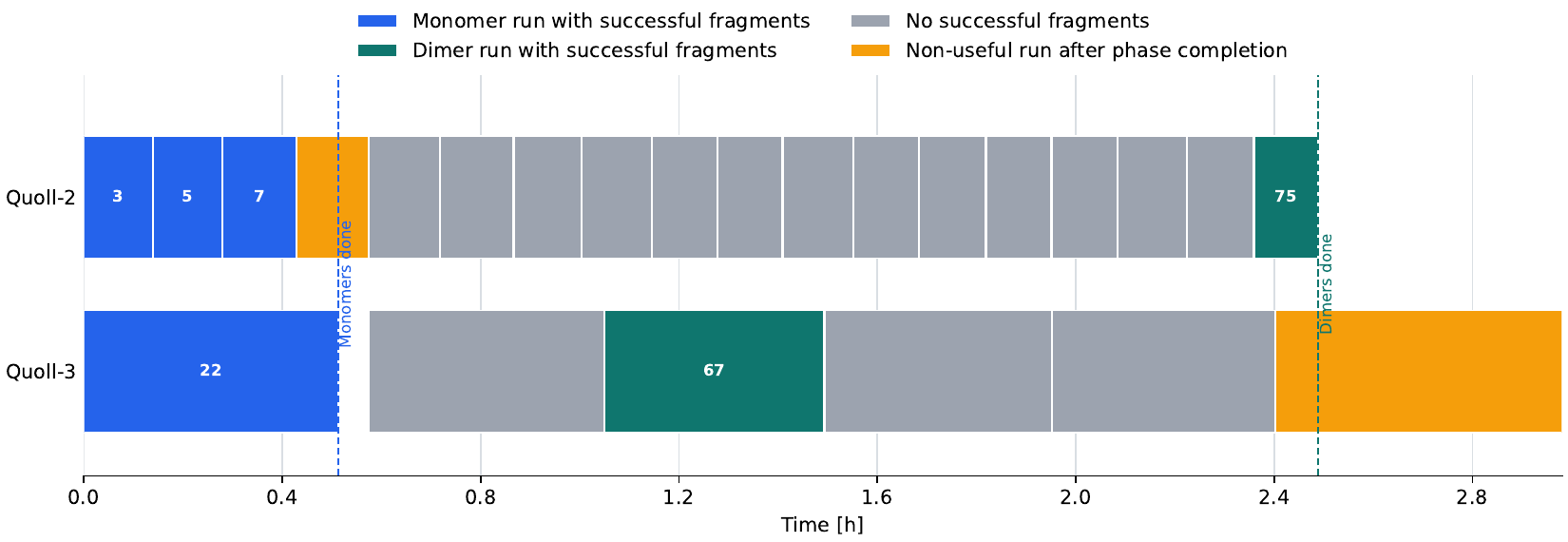}
\caption{\label{fig:frag_parallel_quoll23_activity} Activity diagram Quoll-2+3 asynchronous run. Each horizontal row corresponds to one processor, and each block in each row represents one circuit execution. Blue and green blocks indicate monomer and dimer executions that produced at least one fragment energy within the target threshold, while gray blocks indicate executions that did not yield a newly converged fragment. The orange blocks occur after the final monomer or dimer fragment in a given phase has already been obtained.}
\end{figure*}

To understand this reduction, Fig.~\ref{fig:frag_parallel_quoll23_activity} shows an activity diagram for the Quoll-2+3 asynchronous run.

Ideally, once one processor finds the last remaining fragment in a phase, any still-running circuit executions on the other processor should be terminated immediately, allowing the workflow either to proceed to the next phase or to finish. However, in the implementation used for these experiments, submitted circuit executions were treated as non-interruptible jobs. As a result, once a circuit execution had been dispatched, the workflow waited for it to complete before advancing to the next phase.

This software limitation is especially visible near the end of the dimer phase in Fig.~\ref{fig:frag_parallel_quoll23_activity}: after Quoll-2 obtains the final dimer fragment, it must remain idle while Quoll-3 completes a longer dimer execution that can no longer contribute to the FMO calculation. Since these orange intervals occur only after the relevant phase has already completed, they do not represent a hardware limitation of the processors or a fundamental limitation of asynchronous fragment parallelism. They instead reflect missing job-cancellation functionality in the prototype execution interface.

For this reason, Fig.~\ref{fig:frag_parallel:speedup} also reports idle-time-corrected rates, shown by the opaque markers. These corrected rates remove the waiting time associated with non-useful executions after completion of the corresponding monomer or dimer phase. With this correction, the average shot rates are projected to be $15.3412$~s$^{-1}$, $7.3445$~s$^{-1}$, and $12.5847$~s$^{-1}$ for the three two-QPU configurations, and $17.6343$~s$^{-1}$ for the full three-QPU configuration. The corresponding parallel efficiencies are $99.31\%$, $86.22\%$, $96.12\%$, and $95.17\%$, respectively. These values are much closer to the near-ideal efficiencies obtained in the shot-parallel workflow. Some loss remains, for example in the Quoll-1+3 configuration, because a circuit execution that is nearly complete when the final fragment is found still represents partially wasted time even after the idealized instant-cancellation correction.

Overall, the asynchronous fragment-parallel results demonstrate a second route to practical throughput improvement in the FMO--QWFS workflow. Unlike shot-parallel execution, this mode does not require different processors to sample the same probability distribution, and therefore naturally accommodates device-specific optimal parameters. This feature may make fragment-parallel execution more robust as the number of heterogeneous NISQ processors increases. The main cost is a possible reduction in parallel efficiency due to load imbalance and non-useful circuit executions near phase boundaries. Nevertheless, after accounting for the software-level cancellation limitation of the present prototype interface, the inferred throughput approaches ideal scaling while maintaining chemical accuracy for all two- and three-QPU FMO calculations.

\section{\label{sec:conclusions}Conclusions}

In this work, we investigated whether the natural parallelism of fragment-based quantum chemistry can be translated into a practical execution advantage on real quantum hardware. To this end, we implemented and tested an FMO--QWFS workflow on three deployed quantum processors and applied it to helium clusters from He to He$_{14}$. The central question was not whether an individual quantum processor can solve a small fragment calculation in isolation, but whether multiple quantum processors can be used together to improve the wall-clock performance of a quantum chemistry workflow while preserving the accuracy required for the final FMO energy evaluation.

Our results show that this is possible in two complementary execution modes. In the synchronous shot-parallel mode, multiple processors were used as a combined sampling resource. By distributing the total number of shots according to the measured device shot rates and combining the resulting SPAM-corrected counts into a single probability distribution, we obtained nearly linear increases in sampling throughput. The full three-QPU configuration reached a parallel efficiency of $93.60\%$, while maintaining chemically accurate to the ideal classical FMO-SCI energies for all helium clusters considered. This demonstrates that shot-parallel quantum execution can provide a direct throughput benefit for sampling-based quantum chemistry workflows.

At the same time, the three-QPU shot-parallel analysis revealed an important near-term constraint. Circuit parameters optimized on one processor could not always be transferred directly to a heterogeneous multi-processor sampling configuration, especially for the more noise-sensitive dimer circuits. However, this difficulty could be mitigated either by assigning device-specific optimized parameters to each processor or by optimizing a single parameter set directly under the full three-QPU shot-parallel sampling procedure. Thus, the limitation is not the use of parallel sampling itself, but the need to account for processor-dependent noise when choosing the circuits to be sampled on current NISQ hardware.

The asynchronous fragment-parallel workflow provided a second route to parallel speedup. In this mode, each processor independently sampled complete fragment circuits using its own optimized parameters, while different fragment energies were evaluated concurrently. This approach naturally avoids the cross-device parameter-transfer issue of shot-parallel sampling. 
All two-QPU and three-QPU fragment-parallel runs reached the prescribed acceptance criterion within the number of sampling iterations, and all assembled FMO energies remained within chemical accuracy. After correcting for non-useful idle time associated with non-interruptible submitted jobs in the prototype execution interface, the three-QPU fragment-parallel run achieved an inferred parallel efficiency of $95.17\%$. These results show that fragment-level task parallelism can also be exploited effectively on real quantum processors.

Taken together, these experiments provide a proof-of-concept demonstration that multiple physical QPUs can be used effectively in parallel to increase the throughput of sampling-dominated workloads arising in fragment-based quantum chemistry. The observed advantage is a workflow-level advantage in sampling throughput and wall-clock efficiency, rather than a claim of computational quantum advantage over classical electronic-structure methods. Nevertheless, it addresses a key systems-level question for quantum chemistry: if a molecular problem decomposes naturally into many quantum subproblems, then multiple quantum processors can be used productively to execute those subproblems in parallel.

Looking forward, this perspective becomes even more important in the fault-tolerant era. Realistic quantum chemistry and materials-science applications are expected to require many repeated quantum calculations, whether for fragments, active spaces, embedded regions, geometries, molecular dynamics snapshots, or statistical samples. A future HPC environment equipped with many fault-tolerant quantum processing units could therefore treat quantum processors not as isolated accelerators, but as a parallel computational resource integrated into large-scale electronic-structure workflows. In such a setting, quantum error correction would suppress the hardware variability that complicates present-day NISQ execution, allowing logical circuits to be distributed across many processors with far greater uniformity and reliability.

The present work is a small-scale realization of that broader model. By combining FMO, QWFS, and parallel execution on multiple real quantum processors, we have shown that distributed quantum chemistry workflows can preserve accuracy while improving throughput. This provides an experimental step toward future HPC+QPU architectures in which massively parallel fault-tolerant quantum resources enable chemically accurate simulations of molecular and materials systems beyond the reach of direct classical treatment.

\section*{Supporting Information}

% Supporting Information numbering
\setcounter{section}{0}
\renewcommand{\thesection}{S\arabic{section}}
\renewcommand{\thesubsection}{\thesection.\arabic{subsection}}

\section{Quoll hardware characterization}
\label{sec:SI:Hardware}

\subsection{Quantum-process tomography and gate fidelities}
\label{sec:SI:tomography}

To validate continuous quantum-circuit execution on the deployed QPUs, standard quantum process tomography protocols were performed for multiple angles of the native single-qubit gates, $R_x$ and $R_y$, and for the native anti-controlled two-qubit CZ gate (ACZ). Please note that the ACZ gate was adopted as the native two-qubit gate because we found it to be less susceptible to state cross-talk noise than the conventional CZ gate implementation. Single-qubit performance was characterized by full process tomography measurements of the $R_x(\pi)$, $R_y(\pi)$, $R_x\left(\frac{\pi}{2}\right)$, and $R_y\left(\frac{\pi}{2}\right)$ gates on each qubit. The resulting SPAM-corrected average gate fidelities\cite{Nielsen2002} were averaged to obtain a single-metric single-qubit (SMSQ) gate process fidelity for each QPU. These SMSQ fidelities, together with the ACZ process fidelities, are reported in Table \ref{tab:gate_fidelities}.

\begin{table}[ht]
\centering
\caption{Summary of the measured average gate fidelities of each deployed Quoll. All results are SPAM-corrected and an average was taken of the result of a complete tomography for an $R_x(\pi)$, $R_y(\pi)$, $R_x\left(\frac{\pi}{2}\right)$, $R_y\left(\frac{\pi}{2}\right)$ gate for each qubit to get the single-metric-single-qubit (SMSQ) gate process fidelity for each QPU. In most cases, each circuit was run with 500 shots and each full measurement set was repeated 5 times to calculate the fidelity and the error. The one exception was the ACZ measurement on Quoll 3, which ran with 250 shots and 10 repeats.}
\label{tab:gate_fidelities}
\begin{tabular}{l|c|c|c}
\toprule
\textbf{Gate} &
\textbf{Quoll 1} &
\textbf{Quoll 2} &
\textbf{Quoll 3} \\
\midrule
$R_x(\pi)$@$q_0$ &
$0.99 \pm 0.03$ &
$0.97 \pm 0.02$ &
$0.91 \pm 0.06$ \\

$R_y(\pi)$@$q_0$ &
$0.95 \pm 0.05$ &
$0.98 \pm 0.01$ &
$0.94 \pm 0.05$ \\

$R_x\left(\frac{\pi}{2}\right)$@$q_0$ &
$0.96 \pm 0.04$ &
$0.98 \pm 0.02$ &
$0.90 \pm 0.02$ \\

$R_y\left(\frac{\pi}{2}\right)$@$q_0$ &
$0.95 \pm 0.02$ &
$0.98 \pm 0.01$ &
$0.93 \pm 0.04$ \\

$R_x(\pi)$@$q_1$ &
$0.95 \pm 0.01$ &
$0.97 \pm 0.02$ &
$0.97 \pm 0.09$ \\

$R_y(\pi)$@$q_1$ &
$0.96 \pm 0.018$ &
$1.00 \pm 0.02$ &
$0.97 \pm 0.04$ \\

$R_x\left(\frac{\pi}{2}\right)$@$q_1$ &
$0.98 \pm 0.03$ &
$0.99 \pm 0.02$ &
$0.98 \pm 0.06$ \\

$R_y\left(\frac{\pi}{2}\right)$@$q_1$ &
$0.98 \pm 0.02$ &
$0.98 \pm 0.01$ &
$1.00 \pm 0.03$ \\

\midrule
SMSQ &
$0.97 \pm 0.01$ &
$0.98 \pm 0.01$ &
$0.95 \pm 0.02$ \\

\midrule
ACZ@$q_0q_1$ &
$0.90 \pm 0.01$ &
$0.96 \pm 0.01$ &
$0.79 \pm 0.05$ \\
\bottomrule
\end{tabular}
\end{table}

\subsection{Gate durations and coherence times}
\label{sec:SI:gate_times}

As an indication of the gate times, the time needed for the native $R(\pi)$ ($R_x(\pi)$ or $R_y(\pi)$) and ACZ gates for each qubit and Quoll are collected in Table \ref{tab:quoll_gate_times} alongside the measured $T_2^*$ coherence times. $T_2^*$ coherence times for each qubit were measured using a Ramsey experiment where the coherence time was extracted by fitting the decay envelope of the Ramsey interference fringes. Furthermore, the combined initialisation and readout times for Quoll-1, Quoll-2 and Quoll-3, respectively, are 24\,ms, 33\,ms and 48\,ms.

\begin{table}[h!]
\centering
\caption{Measured Quoll gate and $T_2^*$ coherence times for the native single qubit rotations $R_(\pi)$ ($R_x(\pi)$ or $R_x(\pi)$) and two qubit anti-CZ (ACZ) gate. All times are reported in units of $10^{-6}$\,s.}
\label{tab:quoll_gate_times}
\begin{tabular}{c|c|c|c|c|c}
\toprule
\textbf{Quoll} & \textbf{$R(\pi)$@$q_0$} & \textbf{$R(\pi)$@$q_1$} & ACZ@$q_0q_1$ & \textbf{$T_2^*$@$q_0$} & \textbf{$T_2^*$@$q_1$}\\
\hline
1 & 96.61  & 82.54  & 3.88 & 3.9$\cdot 10^3$ & 7.5$\cdot 10^3$ \\
2 & 136.15 & 100.75 & 3.10 & 1.1$\cdot 10^3$ & 8.1$\cdot 10^3$ \\
3 & 141.74 & 73.37  & 4.00 & 5.5$\cdot 10^3$ & 8.6$\cdot 10^3$ \\
\bottomrule
\end{tabular}
\end{table}

\begin{acknowledgments}

The authors gratefully acknowledge Giuseppe Gebbia, Bryce Henson, James Keaveney, David Keightley, Waseem Kamleh, Lachlan Mackie, Andrea Tabacchini, Zoltán Vörös, and Rishabh Yadav for their contributions and support. The authors also thank all current and former members of Quantum Brilliance whose work contributed to the development of the technologies and capabilities underpinning this work, in particular the QDF, IQC and QSS teams.
This work was supported in part by funding from the Australian Capital Territory (ACT) Government under the Priority Investment Program (Quantum Prototyping Export Program).
L.W.B., L.C., S.I., and D.C. acknowledge support from the Laboratory Directed Research and Development Program of Oak Ridge National Laboratory, managed by UT-Battelle, LLC, for the US Department of Energy. 
This research used resources of the Oak Ridge Leadership Computing Facility and the Compute and Data Environment for Science (CADES) at Oak Ridge National Laboratory, which is supported by the Office of Science of the U.S. Department of Energy under Contract No. DE-AC05-00OR22725.

\end{acknowledgments}

\bibliography{bibliography}

\end{document}